\documentclass[journal]{IEEEtran}
\usepackage{cite}
\usepackage{subfigure}
\ifCLASSINFOpdf
\usepackage[pdftex]{graphicx}
  \graphicspath{{../pdf/}{../jpeg/}}
  \DeclareGraphicsExtensions{.pdf,.jpeg,.png}
\else
  \usepackage[dvips]{graphicx}
  \graphicspath{{../eps/}}
  \DeclareGraphicsExtensions{.eps}
\fi
\usepackage{amsmath}
\usepackage{cases}
\begin{document}
%
% paper title
% Titles are generally capitalized except for words such as a, an, and, as,
% at, but, by, for, in, nor, of, on, or, the, to and up, which are usually
% not capitalized unless they are the first or last word of the title.
% Linebreaks \\ can be used within to get better formatting as desired.
% Do not put math or special symbols in the title.
\title{Privacy-Preserving Probabilistic Forecasting for Temporal-spatial Correlated Wind Farms}
%
%
% author names and IEEE memberships
% note positions of commas and nonbreaking spaces ( ~ ) LaTeX will not break
% a structure at a ~ so this keeps an author's name from being broken across
% two lines.
% use \thanks{} to gain access to the first footnote area
% a separate \thanks must be used for each paragraph as LaTeX2e's \thanks
% was not built to handle multiple paragraphs
%

\author{Mengshuo~Jia,~\IEEEmembership{Student~Member,~IEEE,}
        Chen~Shen,~\IEEEmembership{Senior~Member,~IEEE,}
        ~Zhiwen~Wang,~\IEEEmembership{Student Member,~IEEE}
        and~Zhitong~Yu
        % <-this % stops a space
% \thanks{This work was supported in part by the Foundation for Innovative Research Groups of the National Natural Science Foundation of China under Grant 51621065 and in part by the Joint Funds of the National Natural Science Foundation of China under Grant U1766206 (Corresponding to Chen Shen).}
% \thanks{M. Jia, C. Shen, and Z. Wang are with the State Key Laboratory of Power Systems, Department of Electrical Engineering, Tsinghua University, Beijing 100084, China (e-mails: jms16@mails.tsinghua.edu.cn, shenchen@mail.tsinghua.edu.cn, wang-zw13@mails.tsinghua.edu.cn).}% <-this % stops a space
% \thanks{Manuscript received April 19, 2005; revised August 26, 2015.}
}

% note the % following the last \IEEEmembership and also \thanks - 
% these prevent an unwanted space from occurring between the last author name
% and the end of the author line. i.e., if you had this:
% 
% \author{....lastname \thanks{...} \thanks{...} }
%                     ^------------^------------^----Do not want these spaces!
%
% a space would be appended to the last name and could cause every name on that
% line to be shifted left slightly. This is one of those "LaTeX things". For
% instance, "\textbf{A} \textbf{B}" will typeset as "A B" not "AB". To get
% "AB" then you have to do: "\textbf{A}\textbf{B}"
% \thanks is no different in this regard, so shield the last } of each \thanks
% that ends a line with a % and do not let a space in before the next \thanks.
% Spaces after \IEEEmembership other than the last one are OK (and needed) as
% you are supposed to have spaces between the names. For what it is worth,
% this is a minor point as most people would not even notice if the said evil
% space somehow managed to creep in.

% The paper headers
\markboth{Journal of \LaTeX\ Class Files,~Vol.~14, No.~8, August~2015}%
{Shell \MakeLowercase{\textit{et al.}}: Bare Demo of IEEEtran.cls for IEEE Journals}
% The only time the second header will appear is for the odd numbered pages
% after the title page when using the twoside option.
% 
% *** Note that you probably will NOT want to include the author's ***
% *** name in the headers of peer review papers.                   ***
% You can use \ifCLASSOPTIONpeerreview for conditional compilation here if
% you desire.

% If you want to put a publisher's ID mark on the page you can do it like
% this:
%\IEEEpubid{0000--0000/00\$00.00~\copyright~2015 IEEE}
% Remember, if you use this you must call \IEEEpubidadjcol in the second
% column for its text to clear the IEEEpubid mark.

% use for special paper notices
%\IEEEspecialpapernotice{(Invited Paper)}

% make the title area
\maketitle

% As a general rule, do not put math, special symbols or citations
% in the abstract or keywords.
\begin{abstract}
Adopting Secure scalar product and Secure sum techniques, we propose a privacy-preserving method to build the joint and conditional probability distribution functions of multiple wind farms' output considering the temporal-spatial correlation. The proposed method can protect the raw data of wind farms (WFs) from disclosure, and are mathematically equivalent to the centralized method which needs to gather the raw data of all WFs. 
\end{abstract}

% Note that keywords are not normally used for peerreview papers.

% For peer review papers, you can put extra information on the cover
% page as needed:
% \ifCLASSOPTIONpeerreview
% \begin{center} \bfseries EDICS Category: 3-BBND \end{center}
% \fi
%
% For peerreview papers, this IEEEtran command inserts a page break and
% creates the second title. It will be ignored for other modes.
\IEEEpeerreviewmaketitle

\section{Introduction}

\IEEEPARstart{T}{o} consider the temporal-spatial correlation of multiple wind farms' output (MWO) in probabilistic wind power forecasting, one can first construct the GMM-based joint PDF of MWO at different time periods, and then directly build the conditional PDF of the output of each wind farm (WF) in the next period with respect to the observations of MWO during the current periods\cite{wang2018risk}.  

The construction of the joint and conditional PDF requires complete observations, each of which gathers all the corresponding MWO data at different time periods. Since every WF can only observe its outputs at different time periods, thus the complete observations are vertically partitioned among all the WFs (vertical partitioning: the attributes are divided across sites and the sites must be joined to obtain complete information on any entity \cite{Secure-sum}). However, for protecting data privacy, WFs with different stakeholders may refuse to share those raw data to compose the complete observations for constructing PDF. To solve this privacy issue, the privacy-preserving distributed method is a feasible alternative. 

For constructing the GMM-based PDF, the expectation-maximization (EM) algorithm is commonly used\cite{EMcommonlyused}. Nevertheless, for privacy-preserving distributed EM algorithm, existed researches mainly focus on horizontally partitioned data (horizontal partitioning: each entity is represented entirely at a single site \cite{Secure-sum}). To the best of our knowledge, rarely has literature addressed to deal with the vertically partitioned data to build GMM. Therefore, based on secure multi-party computational (SMC) method\cite{CliftonTool}, this letter proposes a privacy-preserving method to build the GMM-based joint and conditional PDF.

\section{Notations}

We first define domain $\Omega = \{1,2,...,M\}$ for $M$ WFs, $\Gamma = \{1,2,...,T\}$ for $T$ periods (normally $T=24$) and $\Upsilon = \{1,2,...,I\}$ for $I$ observations. Let $\boldsymbol{y}_{m,t}$ denote the random variable of the output for the \textit{m}-th WF at the \textit{t}-th period, where $m \in \Omega$ and $t \in \Gamma$. Then We aim to construct the joint PDF of $\boldsymbol{\rm Y}=\{\boldsymbol{y}_{m,t}|m \in \Omega; t \in \Gamma\}$. The $I$ observations of $\boldsymbol{\rm Y}$ are represented by $\boldsymbol{y}^i = \{y_{m,t}^i|m \in \Omega; t \in \Gamma\}$ ($i \in \Upsilon$). To obtain a complete $\boldsymbol{y}^i$, the corresponding observations of all WFs must be gathered together.

We utilize GMM to build the joint PDF. GMM is a parametric model represented by a convex combination of $J$ multivariate Gaussian distribution functions. We define domain $\Lambda = \{1,2,...,J\}$, then the parameter set of GMM is defined as $\boldsymbol{\rm \theta}=\{w_j,\boldsymbol{\mu}_j,\boldsymbol{\Sigma}_j|j \in \Lambda \}$. The GMM-based joint PDF of $\boldsymbol{\rm Y}$ is given as follows:

\begin{equation}
	\label{joint pdf}
	f({\rm \textbf{Y}};\boldsymbol{\theta})=\sum_{j=1}^J w_j 
	\mathcal N({\rm \textbf{Y}};\boldsymbol{\mu}_j,\boldsymbol{\Sigma}_j)
\end{equation}

\noindent where $w_j$ is the weight coefficient, and
$\mathcal N({\rm \textbf{Y}};\boldsymbol{\mu}_j,\boldsymbol{\Sigma}_j)$ is the \textit{j}-th multivariate Gaussian distribution function with mean vector $\boldsymbol{\mu}_j$ and covariance matrix $\boldsymbol{\Sigma}_j$. The precision matrix is defined as $\boldsymbol{\Phi}_j = (\boldsymbol{\Sigma}_j)^{-1} $. The elements of $\boldsymbol{\mu}_j$ are represented by $\mu_{j,m,t}$ ($m \in \Omega;t \in \Gamma$). The diagonal elements of $\boldsymbol{\Sigma}_j$ or $\boldsymbol{\Phi}_j$ are represented by $\sigma_{j,(m,t),(m,t)}$ or $\phi_{j,(m,t),(m,t)}$ ($m \in \Omega;t \in \Gamma$), and the non diagnoal elements by $\sigma_{j,(m,t),(n,v)}$ or $\phi_{j,(m,t),(n,v)}$ ($m,n \in \Omega;t,v \in \Gamma$).

\section{Construction of The Joint PDF}

To obtain the joint PDF in (\ref{joint pdf}), the key lies in estimating the $\boldsymbol{\rm \theta}$ of GMM. We utilize the EM algorithm to fulfill the estimation. This algorithm is consist of E-step and M-step \cite{EMcommonlyused}. For the \textit{k}-th iteration of the \textit{j}-th Gaussian component, the E-step is given in (\ref{E-step}) and M-step in (\ref{M-step}).

\begin{equation}
	\label{E-step}
	Q_{j}^{i,k+1} = \frac{w_j^k \mathcal N(\boldsymbol{y}^i;\boldsymbol{\mu}_j^k,\boldsymbol{\Sigma}_j^k)}{\sum_{l=1}^J w_l^k N(\boldsymbol{y}^i;\boldsymbol{\mu}_l^k,\boldsymbol{\Sigma}_l^k) } \ ,i \in \Upsilon
\end{equation} 

\begin{subequations}
    \label{M-step}
	\begin{align}
	& w_j^{k+1} = \frac{1}{I} \sum_{i=1}^I Q_j^{i,k+1} \\
	& \boldsymbol{\mu}_j^{k+1} = \frac{\sum_{i=1}^I Q_j^{i,k+1} \boldsymbol{y}^i}{\sum_{i=1}^I Q_j^{i,k+1}}  \\
	& \boldsymbol{\Sigma}_j^{k+1} = \frac{\sum_{i=1}^I Q_j^{i,k+1} (\boldsymbol{y}^i-\boldsymbol{\mu}_j^k)(\boldsymbol{y}^i-\boldsymbol{\mu}_j^k)'}{\sum_{i=1}^I Q_j^{i,k+1}}
	\end{align}
\end{subequations}

Both the two steps require $\boldsymbol{y}^i$ ($i \in \Upsilon$) for calculation. To protect data privacy, we propose a privacy-preserving distributed EM (PDEM) algorithm to handle this privacy issue. The privacy preservation is defined as: the communication data between WFs cannot divulge the raw data.

\subsection{Private E-step}
In the E-step, we assume that all WFs have acquired the $\boldsymbol{\rm \theta}^k=\{w_j^k,\boldsymbol{\mu}_j^k,\boldsymbol{\Sigma}_j^k|j \in \Lambda \}$ updated in the (\textit{k}-1)-th iteration. The aim of the private E-step is to make sure that every WF is able to calculate (\ref{E-step}) without revealing raw data. The essence of (\ref{E-step}) lies in the calculation of the Gaussian component:

\begin{equation}
	\label{Gaussian Component}
	\mathcal N(\boldsymbol{y}^i;\boldsymbol{\mu}_j^k, \boldsymbol{\Sigma}_j^k)  
	=  \frac{exp[-\frac{1}{2} (\boldsymbol{y}^i-\boldsymbol{\mu}_j^k) \boldsymbol{\Phi}_j^k (\boldsymbol{y}^i-\boldsymbol{\mu}_j^k)']}{\sqrt{(2\pi)^{M\times T}|\boldsymbol{\Sigma}_j^k|}} 
\end{equation}
 
\noindent where raw data $\boldsymbol{y}^i$ is only required in the  exponential item $\boldsymbol{g}(\boldsymbol{y}^i) = (\boldsymbol{y}^i-\boldsymbol{\mu}_j^k) \boldsymbol{\Phi}_j^k (\boldsymbol{y}^i-\boldsymbol{\mu}_j^k)'$. We further reorganize $\boldsymbol{g}(\boldsymbol{y}^i)$ into (\ref{private E-step}):

\begin{subequations}
	\label{private E-step}
	\begin{align}	
	& \boldsymbol{g}(\boldsymbol{y}^i) = \sum_{n=1}^M 
	 S_{j,n}^{i,k} - \sum_{n=1}^M \sum_{v=1}^T \mu_{j,n,v}^k \cdot D_{j,n,v}^{i,k} \\
	& S_{j,n}^{i,k} = \sum_{v=1}^T y_{n,v}^i \cdot D_{j,n,v}^{i,k} \ ,n \in \Omega \\
	& D_{j,n,v}^{j,k} =  \sum_{m=1}^M C_{j,m}^{i,k}-H_{n,v}^k \ ,n \in \Omega,v \in \Gamma \\
	& C_{j,m}^{i,k} = \sum_{t=1}^T  y_{m,t}^i \cdot \phi_{j,(m,t),(n,v)}^k \ ,m \in \Omega \\
	& H_{n,v}^k = \sum_{m=1}^M \sum_{t=1}^T \mu_{j,m,t}^k \cdot \phi_{j,(m,t),(n,v)}^k \ ,n \in \Omega,v \in \Gamma
	\end{align}
\end{subequations}

\noindent where (\ref{private E-step}d) and (\ref{private E-step}e) can be calculated by each WF. For calculating (\ref{private E-step}c), each WF has to gather the results of (\ref{private E-step}d) computed by other WFs. Since the results of (\ref{private E-step}d) doesn't reveal the raw data, thus these value calculated by other WFs can be shared. Thereafter, the (\ref{private E-step}b) can be obtained by each WF. For (\ref{private E-step}a), each WF also has to gather the results of (\ref{private E-step}b) of all WFs. Similarly, $S_{j,n}^{i,k}$ in (\ref{private E-step}b) doesn't reveal any raw data, thus this value can also be shared to each WF to calculate (\ref{private E-step}a). Then the Gaussian component in (\ref{Gaussian Component}) is obtainable by every WF. Finally, each WF is able to accurately complete the calculation of the E-step in (\ref{E-step}) by the value of Gaussian component in (\ref{Gaussian Component}) without revealing any raw data. 

\subsection{Private M-step}

After the private E-step, each WF possesses the value of $Q_{j}^{i,k+1}$ ($j \in \Lambda;i \in \Upsilon$). Therefore, every WF is able to compute (\ref{M-step}a) directly. However, $\boldsymbol{y}^i$ is required in (\ref{M-step}b) and (\ref{M-step}c). To avoid revealing raw data, we further reorganize these equations by rearranging the elements of $\boldsymbol{\mu}_j$ and $\boldsymbol{\Sigma}_j$ into (\ref{private M-step mu}) and (\ref{private M-step sigma}), where $m$,$n \in \Omega$ and $t$,$v \in \Gamma$. 

Equation (\ref{private M-step mu}) and (\ref{private M-step sigma}a) for all $T$ time period are obtainable by each WF, and no any WF needs to reveal raw data. Thus, values obtained by (\ref{private M-step mu}) and (\ref{private M-step sigma}a) can be shared among  WFs to compose a complete $\boldsymbol{\mu}_j^k$ and all diagonal elements of $\boldsymbol{\Sigma}_j^k$. 

For (\ref{private M-step sigma}b), the raw data of the $m$-th WF at the $t$-th time period and the $n$-th WF at the $v$-th time period are needed to calculate a scalar product $s_{j,(m,t),(n,v)}^{k+1}$ in (\ref{scalar product}).

\begin{equation}
	\label{private M-step mu}
	\mu_{j,m,t}^{k+1} = \frac{\sum_{i=1}^I Q_j^{i,k+1}y_{m,t}^i}{\sum_{i=1}^I Q_j^{i,k+1}} 
\end{equation}

\begin{subequations}
	\label{private M-step sigma}
	\begin{align}
	& \sigma_{j,(m,t),(m,t)}^{k+1} = \frac{\sum_{i=1}^I Q_j^{i,k+1}(y_{m,t}^i \!-\! \mu_{j,m,t}^k)^2}{\sum_{i=1}^I Q_j^{i,k+1}} \\
	& \sigma_{j,(m,t),(n,v)}^{k+1} \!=\! \frac{\sum_{i=1}^I \!Q_j^{i,k\!+\!1} y_{m,t}^i y_{n,v}^i}{\sum_{i=1}^I Q_j^{i,k+1}} \!-\!\mu_{j,m,t}^k \mu_{j,n,v}^k  
	\end{align}
\end{subequations}

\begin{equation}
	\label{scalar product}
	\begin{aligned}	
	& \boldsymbol{s}_{j,(m,t),(n,v)}^{k+1} = \sum_{i=1}^I Q_j^{i,k + 1} y_{m,t}^i y_{n,v}^i = \boldsymbol{X}_{j,m,t}^{k+1} \cdot \boldsymbol{y}_{n,v} \\
	& \boldsymbol{X}_{j,m,t}^{k+1} = \left[ Q_j^{1,k+1}y_{m,t}^1 \cdots Q_j^{i,k+1}y_{m,t}^i \cdots Q_j^{I,k+1}y_{m,t}^I \right]' \\
	& \boldsymbol{y}_{n,v} = \left[ y_{n,v}^1 \cdots y_{n,v}^i \cdots y_{n,v}^I \right]'
	\end{aligned}
\end{equation}

Since all WFs possess the value of $Q_{j}^{i,k+1}$ ($j=1,...,J;i=1,...,I$), thus knowing both $\boldsymbol{X}_{j,m,t}^{k+1}$ and $\boldsymbol{y}_{n,v}$ means knowing all the raw data. To protect the data privacy, we utilize the secure scalar product (SSP) technique, which can securely compute the scalar product of two vectors, to calculate (\ref{scalar product}). The calculation process of the SSP technique is summarized as follows \cite{CliftonTool}:

\begin{enumerate}
\item Both the $m$-th and $n$-th WF choose a same random $I \times I/2 $ matrix $U$. 
\item The $m$-th WF generates a random $I \times 1$ vector $R$, and send $\boldsymbol{s}_{m} = U\times R + \boldsymbol{X}_{j,m,t}^{k+1}$ to the $n$-th WF.
\item The $n$-th WF calculates the scalar product $\boldsymbol{s}_{n,1} = \boldsymbol{s}_{m} \cdot \boldsymbol{y}_{n,v}$, and also calculates $\boldsymbol{s}_{n,2} = U'\times \boldsymbol{y}_{n,v}$. Then the $n$-th WF send the $\boldsymbol{s}_{n,1}$ and $\boldsymbol{s}_{n,2}$ to the $m$-th WF.
\item The $m$-th WF finally calculates the scalar product through $\boldsymbol{s}_{j,(m,t),(n,v)}^{k+1} = \boldsymbol{s}_{n,1}- \boldsymbol{s}_{n,2} \cdot R$, and then send it to the $n$-th WF.
\end{enumerate}

Through the SSP technique, both the $m$-th and $n$-th WF can acquire the scalar product $s_{j,(m,t),(n,v)}^{k+1}$ without revealing any raw data. Then (\ref{private M-step sigma}b) can be computed by the $m$-th and $n$-th WF ($m$,$n \in \Omega$). Eventually, through sharing (\ref{private M-step mu}) and (\ref{private M-step sigma}a), and utilizing SSP technique, every WF is able to accurately calculate the M-step with the protection of data privacy.

\section{Construction of The Conditional PDF}

Our aim is to construct the conditional PDF of $\boldsymbol{y}_{m,t}$ for the given current outputs of all WFs. Let $v_0$ denote the index of the current time period, then the current outputs is represented by $\boldsymbol{y}_{v_0} = \{y_{m,v_0}|m \in \Omega\}$. Obviously, if $t=v_0+1$, the conditional PDF of $\boldsymbol{y}_{m,t}$ can be viewed as the predictive PDF of the $m$-th WF's output at the next period based on the current outputs of all WFs. 

Once the joint PDF in (\ref{joint pdf}) is built via the PDEM algorithm, the conditional PDF can be constructed:

\begin{equation}
	\label{conditional PDF}
	f(\boldsymbol{y}_{m,t}|\boldsymbol{y}_{v_0}) = \sum_{j=1}^J w_{j,m,t}^c 
	\mathcal N(\boldsymbol{y}_{m,t};\mu_{j,m,t}^c,\Sigma_{j,m,t}^c)
\end{equation}

\noindent where the parameters of the conditional PDF can be specified via (\ref{conditional PDF parameter}):

\begin{subequations}
	\label{conditional PDF parameter}
	\begin{align}
	& w_{j,m,t}^c = \frac{w_j \mathcal N(\boldsymbol{y}_{v_0};\boldsymbol{\mu}_{j,{v_0}}, \boldsymbol{\Sigma}_{j,{v_0}})}{\sum_{l=1}^J w_l \mathcal N(\boldsymbol{y}_{v_0};\boldsymbol{\mu}_{l,{v_0}}, \boldsymbol{\Sigma}_{l,{v_0}})} \\
	& \mu_{j,m,t}^c = \mu_{j,m,t} + \boldsymbol{\Sigma}_{j,{v_0}}^{m,t}(\boldsymbol{\Sigma}_{j,{v_0}})^{-1}(\boldsymbol{y}_{v_0} - \boldsymbol{\mu}_{j,{v_0}}) \\
	& \sigma_{j,m,t}^c = \sigma_{j,(m,t),(m,t)} - \boldsymbol{\Sigma}_{j,{v_0}}^{m,t}(\boldsymbol{\Sigma}_{j,{v_0}})^{-1}(\boldsymbol{\Sigma}_{j,{v_0}}^{m,t})'
	\end{align}
\end{subequations}

\noindent where $\boldsymbol{\mu}_{j,{v_0}}$,  $\boldsymbol{\Sigma}_{j,{v_0}}$ and  $\boldsymbol{\Sigma}_{j,{v_0}}^{m,t}$ are given as follows: 

\begin{equation*}
	\begin{aligned}
		& \!\!\! \boldsymbol{\mu}_{j,{v_0}} = \left[ \mu_{j,1,{v_0}} \cdots \mu_{j,n,{v_0}} \cdots \mu_{j,M,{v_0}} \right] \\
		& \!\!\! \boldsymbol{\Sigma}_{j,{v_0}} = 
			\begin{bmatrix}
				\sigma_{j,(1,{v_0}),(1,{v_0})}  & \cdots\ & \sigma_{j,(1,{v_0}),(M,{v_0})}\\
				\vdots & \ddots  & \vdots  \\
				\sigma_{j,(M,{v_0}),(1,{v_0})}  & \cdots\ & \sigma_{j,(M,{v_0}),(M,{v_0})}
			\end{bmatrix} \\
	    & \!\!\! \boldsymbol{\Sigma}_{j,{v_0}}^{m,t} \!=\! \left[ \sigma_{j,(m,t),(1,{v_0})} \! \cdots \sigma_{j,(m,t),(n,{v_0})} \! \cdots \sigma_{j,(m,t),(M,{v_0})} \right] 
	\end{aligned}
\end{equation*}

Apparently, each WF can compute (\ref{conditional PDF parameter}c) directly with the $\boldsymbol{\rm \theta}$ of the joint PDF. However, to calculate (\ref{conditional PDF parameter}a) and (\ref{conditional PDF parameter}b) needs $\boldsymbol{y}_{v_0}$, which is consist of raw data. To avoid revealing any data privacy, we further reorganize (\ref{conditional PDF parameter}a) into (\ref{private conditional w}) and (\ref{conditional PDF parameter}b) into (\ref{private conditional mu}) . Note that the calculation of (\ref{conditional PDF parameter}a) is similar to the calculation of (\ref{E-step}), thus the reorganization of (\ref{conditional PDF parameter}a) is similar to that of (\ref{E-step}). Due to limited space, we only details the computation parts of (\ref{conditional PDF parameter}a) that have data privacy preserving problem, which is defined as $S_{j,v}^c$ ($v \in \Gamma$) and $C_{j,n,v}^c$ ($n \in \Omega,v \in \Gamma$).

\begin{subequations}
	\label{private conditional w}
	\begin{align}
		& S_{j,v}^c = \sum_{n=1}^M y_{n,v} \cdot D_{j,n,v}^c \\
		& D_{j,n,v}^c = C_{j,n,v}^c - \sum_{l=1}^M \mu_{j,l,v} \phi_{j,(l,v),(n,v)} \\
		& C_{j,n,v}^c = \sum_{l=1}^M y_{l,v} \cdot \phi_{j,(l,v),(n,v)} 
	\end{align}
\end{subequations}

\begin{equation}
	\label{private conditional mu}
	\begin{aligned}
		\mu_{j,m,t}^c = & \mu_{j,m,t} + \sigma_{j,(m,t),(m,v)} \sigma_{j,(m,v),(m,v)}^{-1} y_{m,v} \\
		& - \sum\nolimits_{n=1}^M \mu_{j,m,t} \sigma_{j,(m,t),(n,v)} \sigma_{j,(n,v),(n,v)}^{-1} \\
		& + \sum\nolimits_{n=1,n \not= m}^M \sigma_{j,(m,t),(n,v)} \sigma_{j,(n,v),(n,v)}^{-1} y_{n,v}
	\end{aligned}
\end{equation}

It can be observed that raw data are involved in the weighted sum in (\ref{private conditional w}a), (\ref{private conditional w}c) and the last item of (\ref{private conditional mu}). To avoid revealing raw data, we utilize secure sum (SS) technique, which can securely compute the weighted sum without sacrificing data privacy. Take  (\ref{private conditional w}a) for example, the details of the SS technique are summarized as follows \cite{CliftonTool}: 

\begin{enumerate}
	\item Assume that the sum of (\ref{private conditional w}a) lies in the range [0, N). N can be set as the sum of the capacity of all the WFs. 
	\item The 1st WF generates a random number $Z$, which is uniformly chosen from [0, N). Then the 1st WF send  $V_1 = \left[ (D_{j,1,v}^c y_{1,v} + Z) \ mod \ N \right]$ to the 2nd WF.
	\item For the remaining WFs ($n=2,...,M-1$), the $n$-th WF sends $V_n = \left[ (D_{j,n,v}^c y_{n,v} + V_{n-1}) \ mod \ N \right]$ to the ($n+1$)-th WF. 
	\item When the 1st WF receives the $V_{M-1}$, this WF can finally compute $S_{j,v}^c = (V_{M-1} - Z) \ mod \ N$. Then the value of $S_{j,v}^c$ will be shared among WFs.
\end{enumerate}

With the SS technique, the weighted sum in (\ref{private conditional w}a), (\ref{private conditional w}c) and (\ref{private conditional mu}) can be computed without revealing any raw data. Then the parameters of the conditional PDF in (\ref{conditional PDF parameter}) are obtainable, so is the conditional PDF. 

It's worth noting that the $m$-th WF doesn't participate in the calculation process of the last item in (\ref{private conditional mu}). The value of this item is calculated by the rest WFs, and only useful for the $m$-th WF. Through this design, we can ensure that each WF only can obtain its own conditional PDF without knowing the conditional PDF of others. 

\section{Discussion}

We define the centralized method as the calculation method which can gather the raw data of all the WFs for constructing PDF. Since both SSP and SS techniques can accurately and safely calculate scalar product and weighted sum without any approximation, the proposed method and the centralized method are mathematically equivalent, thus the constructed PDFs of the two method are exactly the same.

The cost of preserving privacy is the increase of  communication traffic. Set $M=10$, $T=24$ and $I=1000$, then in the entire calculation process of the two method, the upstream and downstream total communication traffic of a WF are given in Table \ref{Communication Traffic Comparison}. Since communications occur in every iteration of PDEM algorithm for every observation, thus there is a significant increase for the communication traffic of the proposed method when compared to the centralized method. However, the total communication traffic is still very small and can be fully satisfied under the current bandwidth conditions.

\begin{table}[h]
\renewcommand{\arraystretch}{1.3}
\caption{Communication Traffic Comparison}
\label{Communication Traffic Comparison}
\centering
\begin{tabular}{c c c}
\hline
\bfseries \  & \bfseries Centralized Method & \bfseries Proposed Method\\
\hline
\bfseries Upstream Traffic & 0.08 Mb & 8.93 Mb\\
\bfseries Downstream Traffic & $1.14\times10^{-4}$ Mb & 27.82 Mb\\
\hline
\end{tabular}
\end{table}

The entire process of the proposed method does not require interaction of the WFs with raw data, thus the data privacy is protected. Meanwhile, the proposed method and the centralized method are mathematically equivalent. The communication traffic of the proposed method has increased,  but the total traffic is still very small and can be satisfied.

\ifCLASSOPTIONcaptionsoff
  \newpage
\fi

% trigger a \newpage just before the given reference
% number - used to balance the columns on the last page
% adjust value as needed - may need to be readjusted if
% the document is modified later
%\IEEEtriggeratref{8}
% The "triggered" command can be changed if desired:
%\IEEEtriggercmd{\enlargethispage{-5in}}

% references section

% can use a bibliography generated by BibTeX as a .bbl file
% BibTeX documentation can be easily obtained at:
% http://mirror.ctan.org/biblio/bibtex/contrib/doc/
% The IEEEtran BibTeX style support page is at:
% http://www.michaelshell.org/tex/ieeetran/bibtex/
\bibliographystyle{IEEEtran}
\bibliography{paper}
\end{document}